\documentclass[english]{iopart}
\usepackage[T1]{fontenc}
\usepackage[latin9]{inputenc}
\usepackage{babel}
\usepackage{graphicx}
\usepackage[unicode=true,pdfusetitle,
 bookmarks=true,bookmarksnumbered=false,bookmarksopen=false,
 breaklinks=false,pdfborder={0 0 1},backref=false,colorlinks=false]
 {hyperref}

\makeatletter

\DeclareRobustCommand{\greektext}{%
  \fontencoding{LGR}\selectfont\def\encodingdefault{LGR}}
\DeclareRobustCommand{\textgreek}[1]{\leavevmode{\greektext #1}}
\DeclareFontEncoding{LGR}{}{}
\DeclareTextSymbol{\~}{LGR}{126}

\usepackage{iopams}
\usepackage{setstack}

\usepackage{babel}
\usepackage{caption}
\makeatother

\begin{document}

\title{HoloHands: Kinect Control of Optical Tweezers.}

\author{C. McDonald, M. McPherson, C. McDougall and D. McGloin.}

\address{Electronic Engineering and Physics Division, University of Dundee,
Dundee, DD1 4HN, United Kingdom.}
\begin{abstract}
The increasing number of applications for holographic manipulation
techniques has sparked the development of more accessible control
interfaces. Here, we describe a holographic optical tweezers experiment
that is controlled by gestures which are detected by a Microsoft Kinect.
We demonstrate that this technique can be used to calibrate the tweezers
using the Stokes Drag method and compare this to automated calibrations.
We also show that multiple particle manipulation can be handled. This
is a promising new line of research for gesture-based control that
could find applications in a wide variety of experimental situations. 
\end{abstract}



\maketitle

\section{Introduction}

The versatility of the optical tweezers technique \cite{Ashkin1970}
makes it ideally suited for use and development by a wide variety
of non-expert users. This is highlighted by the increase in the number
of commercial optical tweezers systems from companies such as JPK,
Elliot Scientific and Thorlabs, in addition to the development of
optical tweezers attachments and technologies, such as holographic
optical tweezers. The methods by which optical tweezers are typically
controlled, via beam steering mirrors, acousto-optical deflectors
or by holographic techniques, lend themselves naturally to computer
control. This in turn allows flexible human-computer interfaces to
be developed. The simplest form of this would be mouse control, where
a user points and clicks at a region on the computer monitor where
they desire a trapping spot.

The idea that optical tweezers is a useful biological tool, but one
that is primarily developed and implemented by physicists, while developing
a fairly niche commercial market, has led to the proliferation of
control techniques, usually aimed at enabling the \textquotedbl{}non-expert
user\textquotedbl{} (usually a misnomer) to control sophisticated
systems in a straightforward way. Developments have included touchscreens
\cite{Grieve2009}, haptic feedback \cite{Pacoret2009,Onda2012},
tracking of finger tips \cite{Whyte2006} and most recently the use
of an iPad \cite{Bowman2011}.

The use of tweezers by non-physicists is, however, fairly widespread
and while the argument that non-experts do not wish to realign lenses
on a daily basis and would rather have things tucked inside a sealed
box is clearly true for the most part, it often is not the case in
many of the modern interdisciplinary environments in which academics
now work. Non-expert users can become experts or will often have optical
physicists or engineers as part of their teams or groups. The development
of such techniques is therefore of technical interest but perhaps
of less practical use, at least until touch screen technology based
on tablets is more heavily integrated into commercial optical instrumentation
- a development that may well not be far off, with the release of
iPad control of tweezers by Boulder Nonlinear Systems.

The use of a system that uses whole body tracking, such as the Kinect
presented in this paper, is therefore perhaps of limited cutting edge
scientific interest - although as it is based on high volume commercial
technology, it is low cost and, with the release of a software development
kit (SDK) by Microsoft, it is relatively straightforward to program
using tools such as Visual Studio. As optical tweezers and beam manipulation
technologies are increasingly found in undergraduate teaching laboratories,
the use of a Kinect offers a fairly low-cost interface to control
much higher-end (i.e. expensive) equipment to allow a range of interdisciplinary
skills to be developed for student learning and engagement. In addition,
we suggest that Kinect control of such devices opens them up to use
in areas such as science centres and exhibitions, as well as in other
forms of community outreach activities.

In this paper we outline the development of a holographic optical
tweezers using a Microsoft Kinect as the control mechanism. We illustrate
some of the basic functionality of the device, discuss the limitations
of our implementation and suggest possible extensions to the work.
We also use the Kinect to make some crude measurements of the trapping
efficiency under such conditions and compare this with trapping where
the controller is not a hand waving about in mid-air.

\section{Experimental Methods}

\begin{figure}
\noindent \begin{centering}
\includegraphics[width=0.75\textwidth]{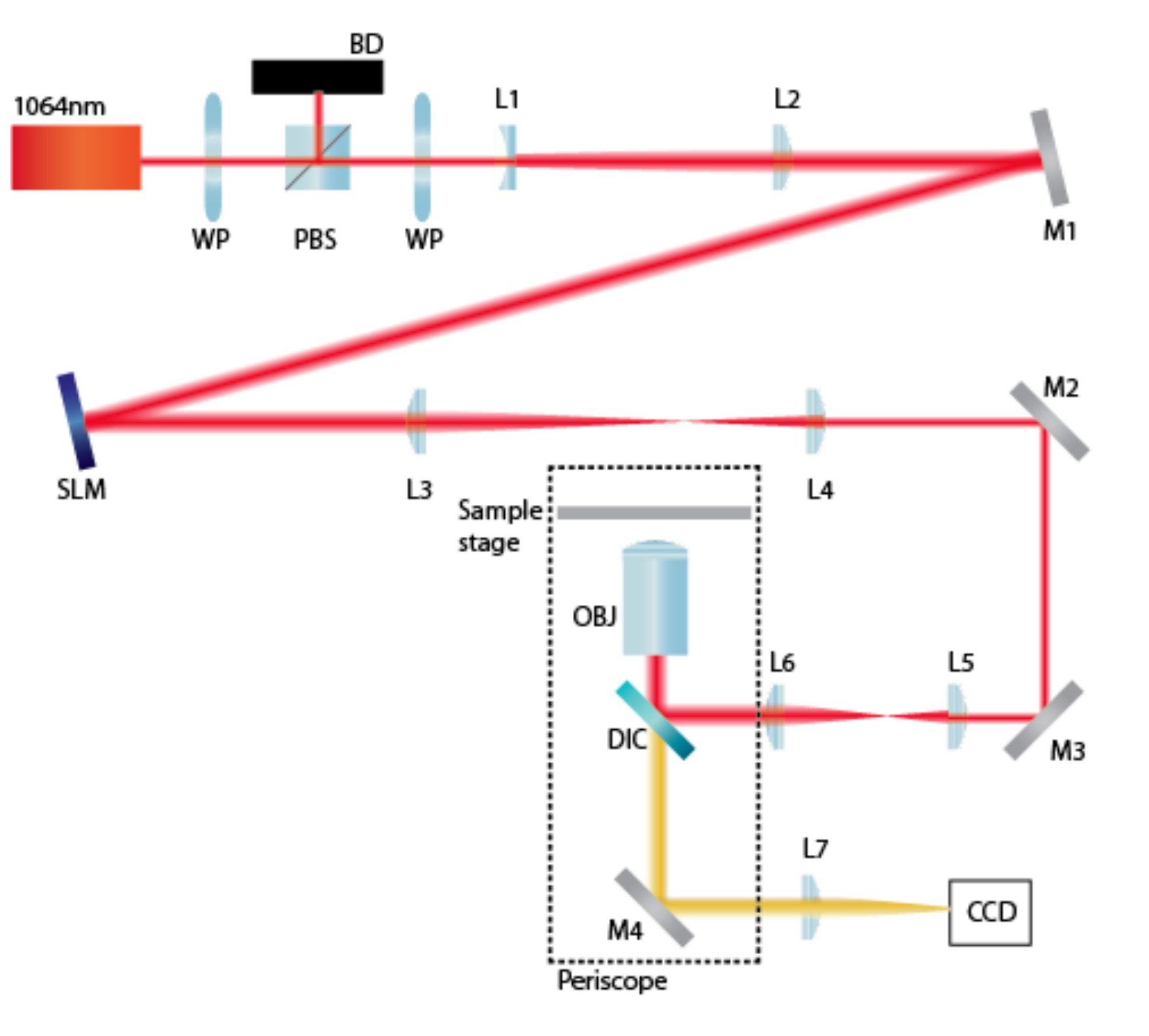}\caption{Kinect controlled holographic optical tweezers setup. The SLM is connected
to the Kinect. WP = half wave plate, PBS = polarising beam splitting
cube, BD = beam dump, DIC = dichroic mirror, OBJ = Nikon E Plan, 100x,
1.25NA microscope objective, CCD = CCD camera, ``L'' denotes ``lens''
with the following focal lengths: L1 = -50mm, L2 = 450mm, L3 = 500m,
L4 = 250mm, L5 = 100mm, L6 = 150mm, L7 = 200mm, M\# = mirror. Not
shown here: Kohler illumination, a zero-order beam block placed between
L3 and L4 and a Newport M562 series manual xyz translation stage with
custom made sample holder.\label{fig:Kinect setup}}

\par\end{centering}

\end{figure}

\noindent The kinect controls a standard holographic optical tweezers
(HOT) setup \cite{Burnham2006}, figure \ref{fig:Kinect setup}. Our
optical source is a 10W (max power) 1070nm fibre laser (IPG Photonics)
expanded to slightly overfill a Holoeye PLUTO spatial light modulator
(SLM) designed for use around 1064nm. Both the SLM and the Microsoft
Kinect were connected to a Windows 7 based PC with an Intel Xeon E31270
processor running at 3.40GHz with 8GB of RAM, which was also used
for the development, and running, of our programs. 

Initial program development work made use of the C\# wrapper for the
OpenNI libraries as Microsoft had not, at the time, released their
official SDK for Kinect for Windows. However, this was subsequently
released during the early stages of our development and a decision
to switch to this implementation was made. This led to more straightforward
coding and gave library functions which were better documented. However,
with the full development system now being used, executed programs
were seen to be a little more sluggish in execution - which could
be worth considering depending on the specific application in mind.
The differences in speed and efficiency were not quantified - we found
the SDK straightforward to use and all the results presented made
use of this implementation of the code.

\begin{figure}
\noindent \begin{centering}
\includegraphics[width=0.75\textwidth]{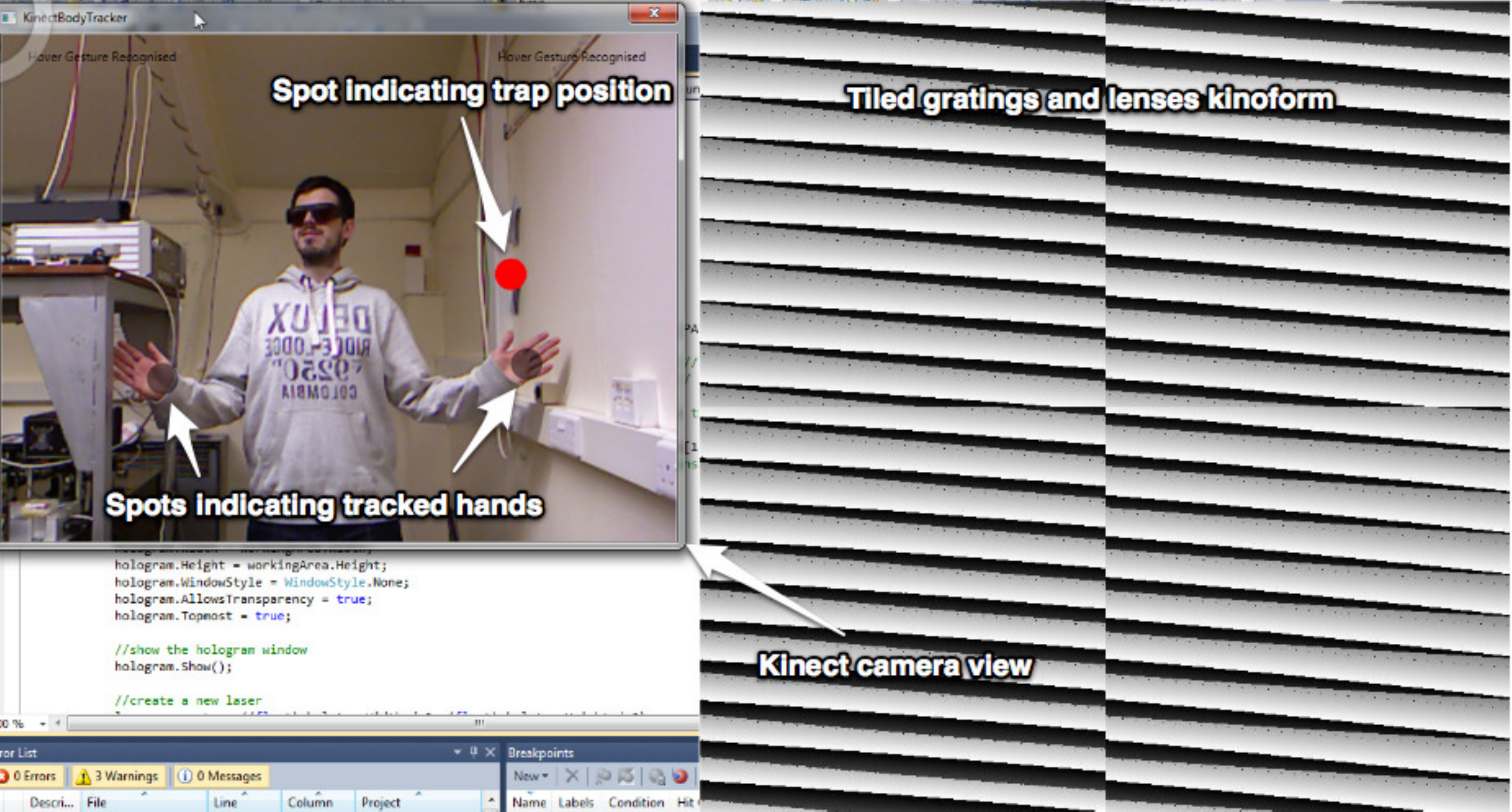} 
\par\end{centering}

\noindent \raggedright{}\caption{Example of the HoloHands program. A video display from the Kinect
camera, shown in the upper right of the figure, allows the user to
view themselves. The red spot on the image corresponds to the position
of the trapping spot and, depending upon the program configuration,
can be controlled by hand tracking or by simple automation of the
SLM. The spot location is then translated to the grating kinoform
that, when applied to the SLM, results in the motion of the trapping
spot and, ultimately, a trapped particle.\label{fig:Matt+Tiles}}
\end{figure}

The program which was developed gives control to the user through
a series of pre-programmed hand gestures. For example, when a free
hand is waved, a new trapping spot is created and can be moved by
that hand. Waving the hand again, when a trapping spot is ``active'',
will delete the spot. A trapping spot can be locked in position through
a ``clicking'' gesture - moving the hand with control of the trapping
spot away from the body and back will ``put down'' the translating
spot. Control of this spot can then be regained by hovering a free
hand above the stationary spot. Through the use of this basic functionality
and these simple gestures, more complex control can be built up: looking
at multiple hand tracking or manipulation of multiple spots, for example.
All of these functions are demonstrated in the HoloHands video, available
in the online supplementary data. The control program translates these
hand gestures into a kinoform, a phase-only hologram. A simple gratings
and lenses \cite{Liesener2000} algorithm was used as the basis for
hologram generation but, when implemented across a 1080x1080 pixel
hologram, we found the program response to be sluggish. Moving to
a tiling strategy, making use of typically four tiles, greatly increased
the speed. Through further optimisation of hologram calculation and
display, it should be possible to increase program efficiency and
speed in order to achieve a far smoother response. Figure \ref{fig:Matt+Tiles}
shows the Kinect camera view and the corresponding tiled kinoform
which would be displayed on the SLM. 

\section{Results and Discussion}

While allowing for a more intuitive control of holographic optical
tweezers, the interface between the Kinect and the optical tweezers
is not the most practical for every day work. For instance, after
a short period of time, use of the arms to position things becomes
tiring. Proper elimination of stray motions detected by the Kinect
can be troublesome. We also found that, although program response
had been increased through tiling, there remained some lag in initiating
functions, such as creating a trapping spot. While we are confident
that these can be improved via better hardware or software implementation,
but it does pose limitations for quantitative work.

\begin{figure}
\noindent \begin{centering}
\includegraphics[width=0.75\textwidth]{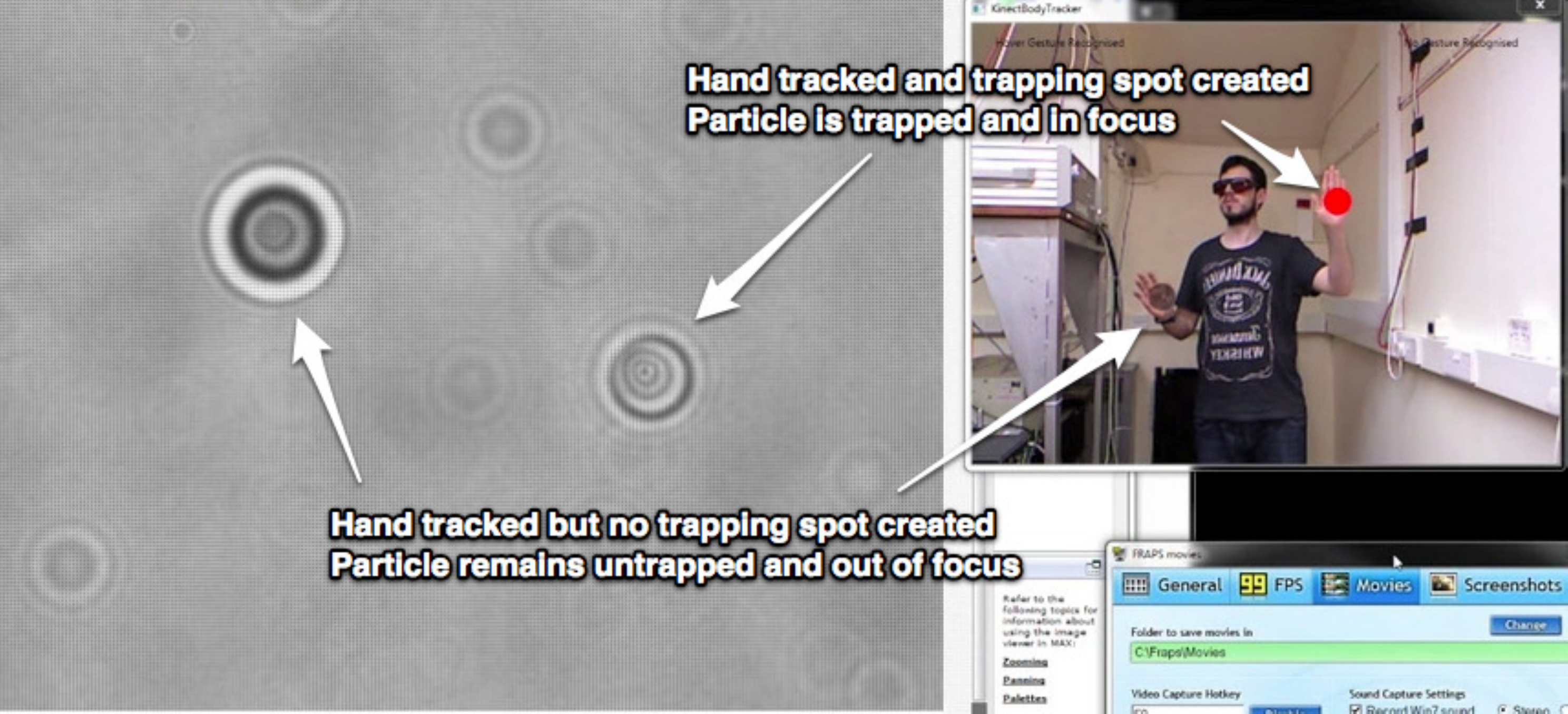} 
\par\end{centering}

\noindent \centering{}\caption{Screenshot of Kinect holographic optical tweezers control. User is
in the right hand screen with the microscope view in the left hand
image. The use of a tracked hand to move a single spot is shown here.
The trapped particle is shown in focus while the untrapped particle
remains out of focus.\label{fig:One Trapped}}
\end{figure}

\begin{figure}
\begin{centering}
\includegraphics[width=0.75\textwidth]{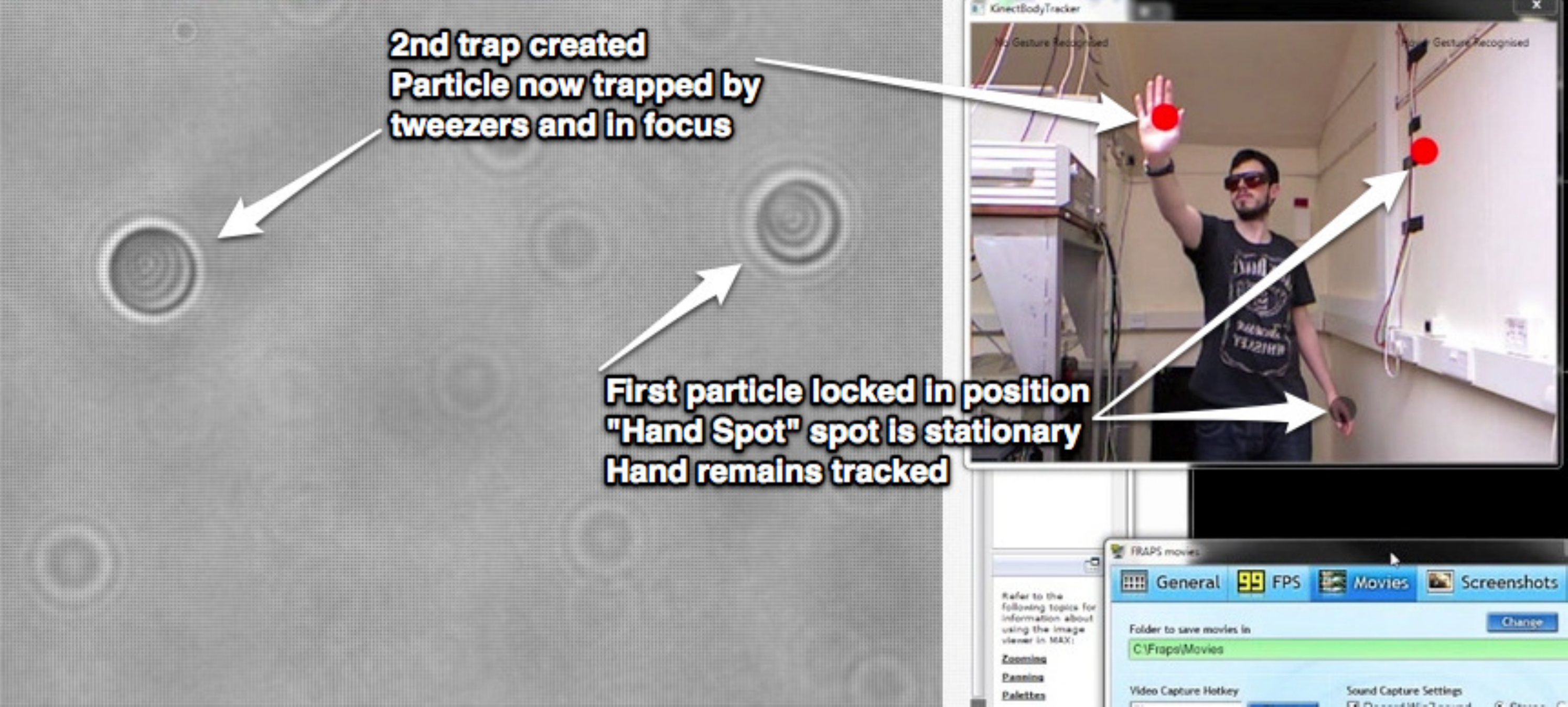} 
\par\end{centering}

\caption{Two trapped particles. The right ``hand'' particle is being manipulated
by the user via the Kinect, while the left ``hand'' particle is
locked in place.\label{fig:Two-trapped-particles.}}
\end{figure}

\begin{figure}
\begin{centering}
\includegraphics[width=0.75\textwidth]{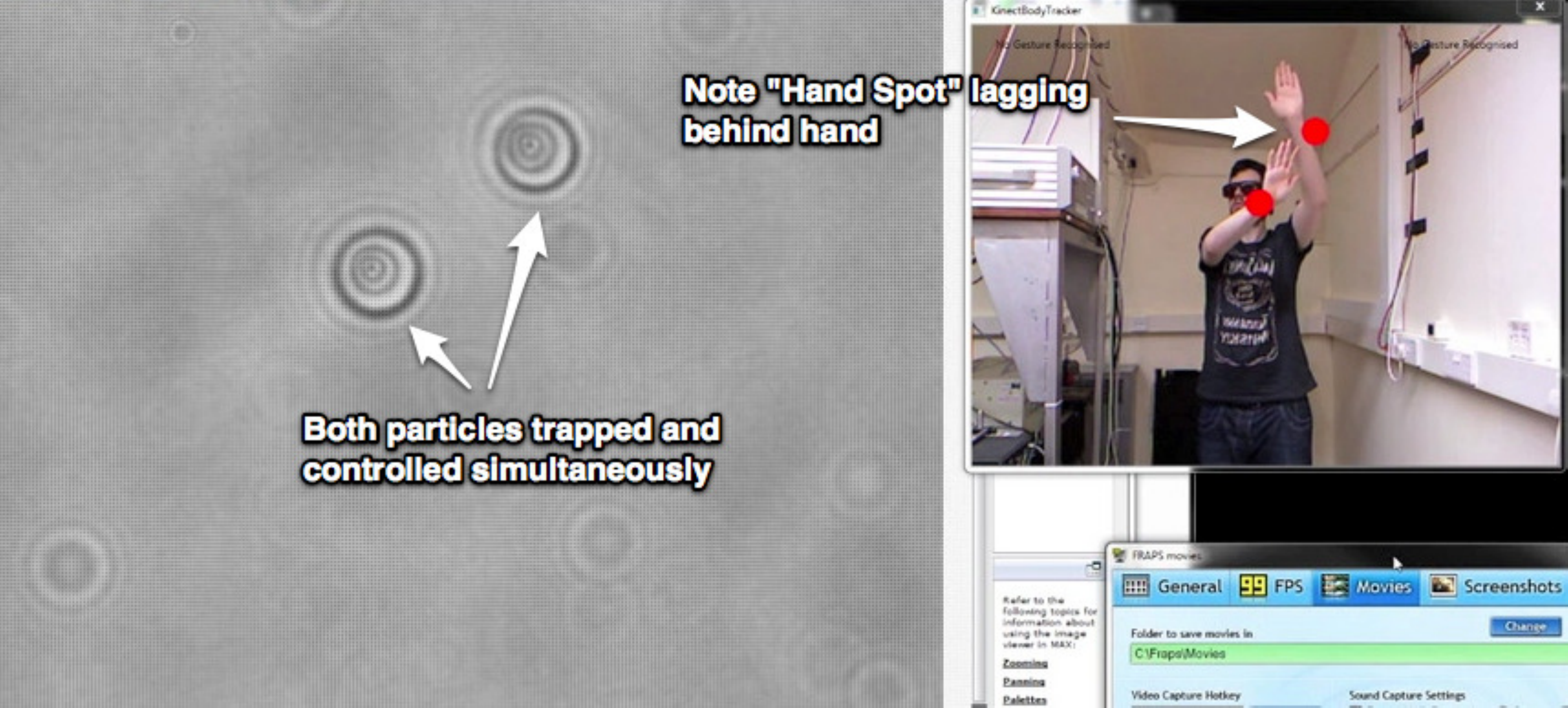} 
\par\end{centering}

\caption{Kinect control of two trapped particles. The lag in movement of the
hand and the tracking position is also highlighted by the spot near
the user's left hand.\label{fig:Swapping Spots}}
\end{figure}

Figures \ref{fig:One Trapped}, \ref{fig:Two-trapped-particles.}
and \ref{fig:Swapping Spots} show the basic functionality of the
Kinect control system. The user is shown in the top right hand corner,
hand tracking is highlighted with the light grey spots and a trapping
spot is represented by the red spot in the user's view. Figure \ref{fig:One Trapped}
shows the user, having initiated a ``wave'' gesture to create a
trapping spot, trapping a single, 4.32\textgreek{m}m diameter, silica
particle with their left hand. Although the right hand is also being
tracked, represented by the light grey spot, the lack of a red spot
indicates that a second trapping spot has not been generated, hence
the untrapped second particle. A second ``wave'' gesture, this time
with the right hand, creates a second trapping spot which can then
be used to trap the second particle, as shown in figure \ref{fig:Two-trapped-particles.}.
Note that the first particle is now locked in position and the ``hand''
spot is now stationary due to the execution of a ``clicking'' gesture.
Finally, multiple spot movement can be demonstrated, as shown in figure
\ref{fig:Swapping Spots}, where two particles orbit one another.
Also shown in figure \ref{fig:Swapping Spots} is an example of the
slight latency in our program concerning the ability of the hand tracker
to keep up with hand movements.

Program speed was improved over time through the introduction of threading
into the program and executing different threads for different sections
of the program, such as generation of the final hologram. The way
in which the bitmaps used to hold and combine kinoforms were manipulated,
using the \texttt{GetPixel} and \texttt{SetPixel} methods, proved
to be a bottleneck in the program. Moving to a solution utilising
the \texttt{LockBits} method to manipulate byte arrays resulted in
a 6-times speed improvement in kinoform calculation. Through further
optimisation of the calculation algorithm, we are confident that more
significant speed gains could be achieved. 

\subsection{Efficiency Measurements}

In order to examine the ability of the Kinect control system to function
as a more robust research grade instrument, a simple measure of the
transverse Q value \cite{Ashkin1992} of the trapping spot was made.
By making use of a Stokes' drag technique, in conjunction with the
expression $F=Q(n_{m}P/c)$, where $F$ and $P$ are, respectively,
the trapping force and power, $n_{m}$ is the suspending medium's
refractive index and $c$ being the speed of light, the trapping efficiency
of the system was determined. Two implementations were compared in
order to determine the effect of user interaction. Initially, a holographic
spot was generated and automatically scanned across the experimental
field of view. This was then repeated but with hand control used to
move the particle, rather than automatically scanning the spot. In
both cases, for a given power, the speed at which the particle fell
out of the trap was measured. Silica spheres of 4.32\textgreek{m}m
diameter were trapped with 55mW of laser power and a Q value of $0.0079\pm0.0002$
was measured for automatic motion, while hand movement control yielded
a Q of $0.0045\pm0.0002$ for the same trapping power. It was experimentally
verified that the light intensity in the trapping spot did not differ
significantly across the experimental field of view. The drop in the
Q value between the two cases was unexpected because the same trapping
spot had been used, and translated in the same direction, in both
cases. We attributed this drop in Q value to the fact that keeping
hand gestures smooth and at a constant velocity for the full experimental
run was often difficult to achieve.

In order to quantify and gain a better understanding of the smoothness
of the hand movements, particle tracking was performed on a number
of videos through the use of ImageJ software \cite{ImageJ}. Figures
\ref{fig:Displacement-vs-Time}a, \ref{fig:Displacement-vs-Time}b
and \ref{fig:Displacement-vs-Time}c show displacement vs time graphs
for three different hand movement videos. In each case, the trapped
particle, the red dot produced by the Kinect tracking, and the user's
hand were tracked. Figure \ref{fig:Displacement-vs-Time}a shows that
there is a close correlation between all three components - in this
case the hand movement was sufficiently smooth to produce a straight
line, with the movement of the red dot and trapped particle following
closely. This is not the case with the video tracked to produce figure
\ref{fig:Displacement-vs-Time}b, where, although the hand movement
is reasonably smooth, the red dot occasionally lags behind the hand
movement, resulting in large jumps in order to ``catch up'' with
the tracked hand. These large jumps by the red dot are then translated
to large jumps by the particle approximately 0.5s later, highlighting
the previously mentioned lag-time in our system. This is more evident
in figure \ref{fig:Displacement-vs-Time}c, where the hand movement
is the least smooth of all three cases. Here it is clear that the
red dot, corresponding to the Kinect tracking, can often experience
a large lag-time and then jump position, in a fraction of a second,
to catch up with the tracked hand. Once translated to the kinoform,
the particle can experience jumps of up to 6\textgreek{m}m in as little
as 0.034s. This could have contributed to the drop in Q value as the
average speed over each experimental run was used in the calculation,
neglecting any instantaneous jump the particle may have experienced
during the measurement. It is worth noting, however, that figures
\ref{fig:Displacement-vs-Time}b and \ref{fig:Displacement-vs-Time}c,
which show these large jumps in position and lag time, correspond
to a particle traveling at roughly twice the speed of the particle
in figure \ref{fig:Displacement-vs-Time}a. This also gives an indication
of the difficulty in performing quantitative measurements with our
system. Qualitative measurements and trial experiments, however, can
be easily performed and non-expert users would certainly be able to
make use of HOTs through our intuitive interface. The minimum trapping
power in our trap was approximately 1mW.

\begin{center}
\begin{figure}
\centering{}\includegraphics[height=0.66\paperheight]{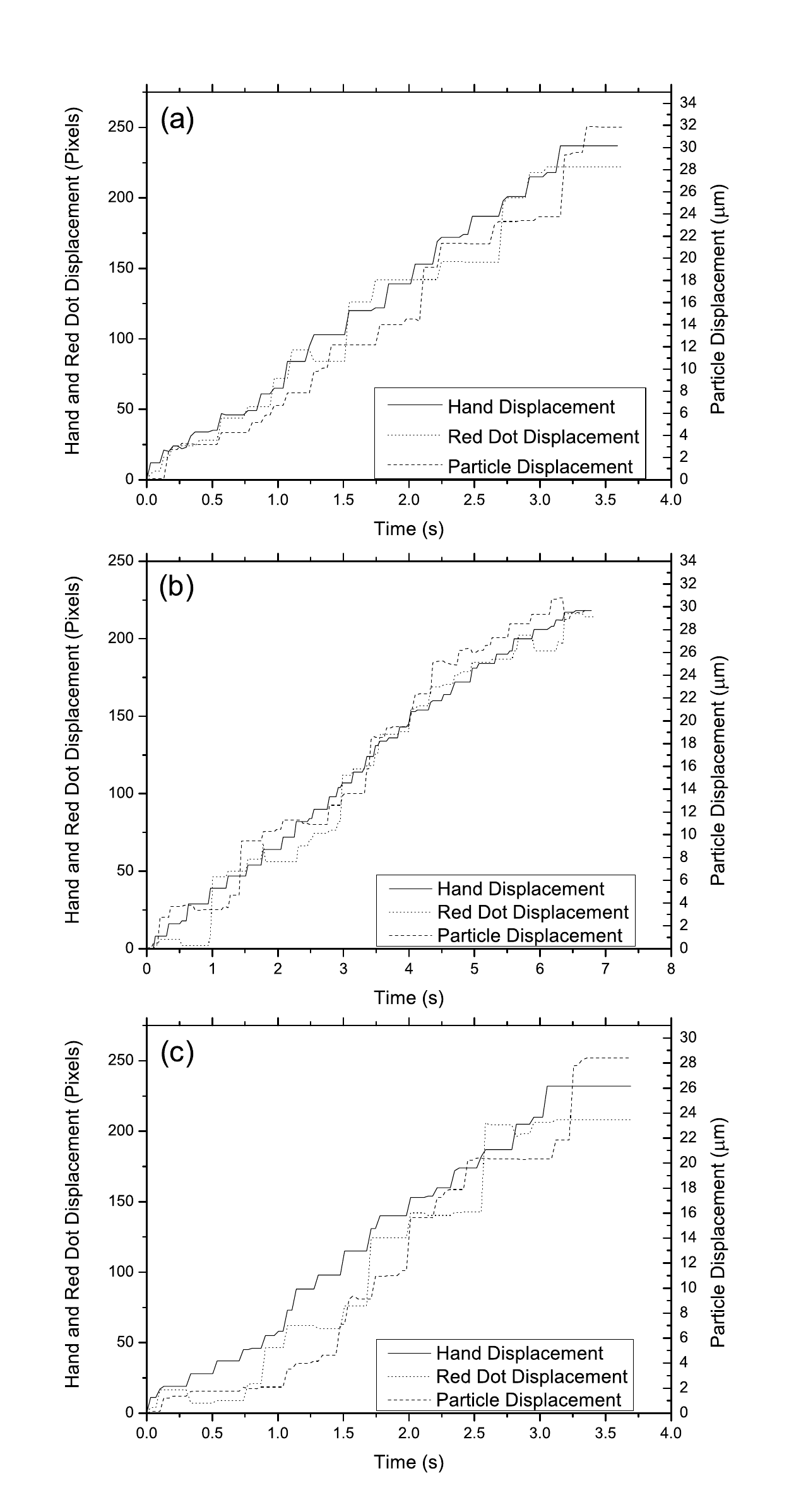}\caption{Displacement vs Time graphs for three separate tracking events. In
each graph, the trapped particle, the red Kinect tracking dot and
the User's hand were tracked.\label{fig:Displacement-vs-Time}}
\end{figure}

\par\end{center}

\section{Future Work}

In this work we have provided a proof of concept of the type of control
which is possible using a full-body interface for a set of optical
tweezers. Although we have made use of an SLM, our Kinect control
system could easily be configured to interface with any other computer-based
tweezers control system, such as an AOD. Sensitivity was an issue
in our experiments in so far as the image size did not scale with
the size of the person being tracked. As such, a person with a lager
arm span could move trapped particles further and with greater precision.
The skeleton detection algorithm could be used to provide some form
of calibration in order to resolve this issue.

On occasion, there were issues with the skeleton detection routine,
causing our tracking software to find skeletons based on inanimate
objects present in the scene. This could cause a set of co-ordinates
to be generated for hand positions which do not exist. Future iterations
of our program would have a formal activation for a second person
entering the scene to act as a secondary control ``mechanism''.

Gesture detection could be improved in order to prevent similar gestures
being confused. This would also allow the straightforward incorporation
of a greater number of gestures, such as those which would create
other kinds of kinoforms for Laguerre-Gaussian or Bessel beams, for
example.

\section{Conclusions}

Our Microsoft Kinect based interface allows for intuitive control
of a set of holographic optical tweezers. Our versatile system clearly
shows the straightforward techniques required for a basic research
grade tweezers system but it lacks the ability to perform quantitative
measurements without some extra sensitivity or smoothing function
to buffer the tracking and particle motion. This makes it challenging
to use for precision work, but will eventually make it suitable for
demonstrations in schools or science centres. High precision work
could, in time, be carried out with new technology currently coming
to market, such as the Leap Motion, which alleges to be 200 times
more accurate than other motion detecting devices and having the ability
to track movements as small as 10\textgreek{m}m \cite{LeapMotion}.
The Kinect itself, coupled with the SDK, offers a cheap computer control
for a variety of experimental systems. The type of experiment described
above would make an ideal undergraduate project or investigation,
with the aim of developing interdisciplinary skills or interdisciplinary
team-working with students from across traditional academic subject
areas.

\ack{}{We thank the UK EPSRC for support, grant EP/H004238/1. DM thanks
the Royal Society for their support in granting the award of a University
Research Fellowship.}

\section*{References}

\bibliographystyle{unsrt}
\bibliography{KinectRef}

\end{document}